\documentclass[preprint,aps,pra,showpacs,floatfix]{revtex4}
\usepackage{amsmath}
\usepackage{times}
\usepackage {euscript}
\usepackage{graphicx}
\usepackage{amsthm}
\usepackage{epsf}
\usepackage{epsfig}
\usepackage{bm}
\renewcommand{\vec}[1]{{\mbox{\boldmath$#1$}}}

\begin{document}
\thispagestyle{empty}
\title{Relativistic calculations of the U$^{91+}(1s)$--U$^{92+}$
collision using the finite basis set of cubic Hermite splines on a
lattice in coordinate space}
\author{G.~B.~Deyneka,$^{1}$ I.~A.~Maltsev,$^{2}$ 
I.~I.~Tupitsyn,$^{2}$
V.~M.~Shabaev,$^{2}$
A.~I.~Bondarev,$^{2}$
Y.~S.~Kozhedub,$^2$
G.~Plunien,$^{3}$
and Th.~St\"ohlker$^{4,5,6}$
}
\affiliation{
$^1$ St. Petersburg State University of Information Technologies,
Mechanics and Optics, Kronverk av. 49, 197101 St. Petersburg,  
Russia \\
$^2$ Department of Physics, St. Petersburg State University,
Ulianovskaya 1, Petrodvorets, 198504 St. Petersburg, Russia
\\
$^3$ Institut f\"ur Theoretische Physik, Technische Universit\"at Dresden,
Mommsenstra{\ss}e 13, D-01062 Dresden, Germany\\
$^4$
GSI Helmholtzzentrum f\"ur Schwerionenforschung GmbH,
Planckstrasse 1, D-64291 Darmstadt, Germany \\
$^5$Helmholtz-Institute Jena, D-07743 Jena, Germany\\
$^6$Institut f\"ur Optik und Quantenelektronik,
Friedrich-Schiller-Universit\"at,
D-07743 Jena, Germany
\vspace{10mm}
}
%
\begin{abstract}
A new method for solving the time-dependent two-center
Dirac equation is developed. The approach is based on the using of the finite
basis of cubic Hermite splines
on a three-dimensional lattice in the coordinate space. 
The relativistic calculations of the excitation and 
charge-transfer probabilities in the low-energy U$^{91+}(1s)$--U$^{92+}$ collisions in two
and three dimensional approaches are performed.
The obtained results are compared with our previous calculations employing
the Dirac-Sturm basis sets  [I.~I. Tupitsyn \textit{et al.}, Phys. Rev. A
\textbf{82}, 042701 (2010)]. 
The role of the negative-energy Dirac spectrum is investigated within
the monopole approximation.
\end{abstract}
\maketitle
\section{INTRODUCTION}
%
Heavy-ion collisions play a very important role in studying relativistic
quantum dynamics of electrons in the presence of strong electromagnetic fields
\cite{Eichler_95,Shabaev:pr:2002,Eichler:pr:2007,Tolstikhina:pu:2012}. 
Such
collisions can also give a unique tool for tests of quantum electrodynamics
at the supercritical fields, provided the projectile energy approaches the Coulomb barrier 
(about 6 MeV/u for the $\rm U$-$\rm U$ collisions) \cite{Greiner_85}.
To date various theoretical methods were developed for calculations of
heavy-ion collisions. Among them are the lattice methods for solving the
time-dependent Dirac equation in the
coordinate
space~\cite{Becker86, Strayer90, Thiel92, Wells92, Wells96,
Pindzola_00, Busic_04} 
and in the momentum space~\cite{Momberger96, Ionescu99}.

In the case of head-on collisions, due to the rotational symmetry
with respect to the internuclear axis, the three-dimensional (3D) process
is easily reduced to the two-dimensional (2D) one.  
Moreover, to simplify the numerical procedure, the 2D approximation
can be applied for the 3D collision with nonzero impact parameter as well. 
This simplification was used in Refs.~\cite{Becker86,Thiel92},
where the calculations were performed by the finite difference method on a
two-dimensional grid.

In Refs.~\cite{Eichler90, Rumrich93,Momberger93, Gail2003} the basis sets of
atomic eigenstates were employed to study heavy-ion
collisions at high energies.
The authors of Refs.~\cite{Kurpick92, Kuprick95} studied various processes in low-energy
ion-atom collisions with the use of relativistic molecular orbitals. 
In works~\cite{Muller_72,Soff_78,Reus_84,Ackad_08} some effects
were
investigated in
so-called monopole approximation, which allows one to reduce the 2D and 3D
two-center Dirac equations to the spherically symmetric one-center radial
equation. The monopole approximation was found to be very useful for studying
processes at short internuclear distances~\cite{Muller_94}. Unfortunately,
this approach as well as its one-center extensions 
beyond the monopole approximation~\cite{Horb11,Mac12}
can not be applied to calculations of charge-transfer processes. 
Recently~\cite{Tup_10,Tup_12,Kozhedub_2013} we developed a method which allows
solving the
time-dependent two-center Dirac equation in the basis of atomic-like
Dirac-Fock-Sturm orbitals. With this method we could calculate the
electron-excitation and charge-transfer probabilities in low-energy ion-ion and
ion-atom collisions.
Despite the diversity of the methods developed, none of them provide
the full relativistic treatment of the quantum dynamics of electrons
in low-energy heavy-ion collisions beyond the monopole approximation.
In particular, it means that with these methods we can not calculate
the charge-transfer probability for the low-energy collision with
the proper account for the dynamics of the occupied negative-energy
states. This problem, which seems especially important for studying
the supercritical regime, remains unsolved even for the simplest
one-electron case. Moreover, the rather successful application
of the method of Ref.~\cite{Tup_10} for calculations of the charge-transfer and 
total
ionization probabilities and its generalization to study the excitation 
and charge
transfer with many-electron systems~\cite{Tup_12, Kozhedub_2013} 
does not guarantee that the
finite basis set representation based on the atomic-like orbitals can 
properly
describe the two-center continuum states.

In the present paper, which should be considered as a continuation of our
previous
investigations~\cite{Tup_10,Tup_12,Kozhedub_2013,Bondarev_2013,Maltsev_2013}, we
work out an alternative
approach to calculations of electron-excitation and charge-transfer
probabilities in low-energy heavy-ion collisions. In this method, the
time-dependent Dirac wave function is expanded in the basis of Hermite cubic
splines at a fixed grid. Such a basis was previously successfully used in the
1D and 2D time-dependent nonrelativistic
calculations~\cite{Deineka_04,Deineka_06}.

The Hermite splines are a special choice of well-known
$B$-splines~\cite{De_Boor_01}. During the last decades
the $B$-splines were successfully applied for solving the
one-center Dirac equation~\cite{Johnson_88,Shabaev:prl:04} as well as the
two-center
nonrelativistic Schr\"odinger~\cite{Artemyev_04} and relativistic
Dirac~\cite{Artemyev_11,Mac12} problems. The Hermite splines have
been used to
obtain accurate solutions of the nonrelativistic
Hartree-Fock~\cite{Morrison_96, Morrison_00} and relativistic Dirac-Fock
equations~\cite{Deineka_96} for diatomic molecules.
An accurate finite element method using cubic Hermite splines was also
recently developed for atomic calculations within the density
functional theory and the Hartree-Fock method~\cite{Ozaki_11}. The convergence
with respect to the total number of the basis functions was investigated for
Hermite splines of different order and it was concluded that the cubic splines
provide an optimum choice with respect to the convergence and the simplicity
of analytic expressions derived for the matrix elements~\cite{Ozaki_11}. The
basis of the cubic Hermite splines is shortly discussed in Sec.~\ref{sec:A2} of
the present paper.

In Sec.~\ref{sec:methods} we describe the procedure of solving the one-electron
time-dependent Dirac equation in the finite basis of the Hermite cubic splines.
The monopole (1D), axially symmetric (2D) and full 3D approaches are
formulated. Basic formulas for the transition amplitudes, including those which
properly account for the negative-energy spectrum contribution, are also given. 
In~Sec.~\ref{sec:results} the results of our calculations of the excitation and
charge-transfer probabilities for the U$^{91+}(1s)$--U$^{92+}$ collision at the
projectile energy $E=6$~MeV/u are presented and compared with the previous
calculations.

Atomic units ($\hbar=e=m=1$) are used throughout the paper.
%
\section{Methods of Calculation }
\label{sec:methods}
\subsection{Time-dependent Dirac equation in a finite basis}
\label{sec:A1}
\begin{figure}
\centering 
\includegraphics[trim=0 0 0 0, clip, width = 0.6\textwidth]
{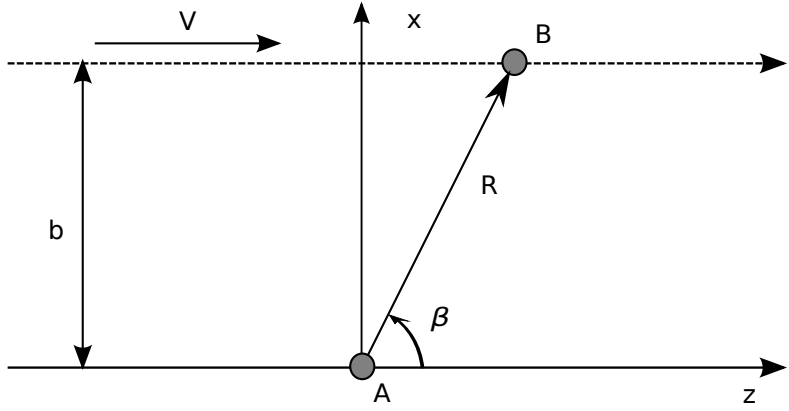}
\caption{The straight-line trajectory of the ion-ion collision. 
The target $A$ is stationary, while the projectile $B$ moves along a straight
line with the velocity $v$. $R$ is the distance between the target $A$ and the
projectile $B$, and $b$ is the impact parameter.
}
\label{fig_geometry}
\end{figure}

%
In our consideration we employ the semiclassical approximation, where the
atomic nuclei are treated as sources of a time-dependent external potential. 
What is more, instead of using the classical (Rutherford) trajectories,
in our calculations we assume that the projectile (U$^{92+}$) moves along a
straight line with a constant velocity, while the position of the target
(U$^{91+}(1s)$) is fixed~(Fig.~\ref{fig_geometry}). 
The electron motion
is described by the time-dependent Dirac equation
\begin{equation}
i\frac{ \, \partial}{\partial t\,} \, \psi(\textbf{r},t) = \hat H  \,
\psi(\textbf{r},t) \,, 
\qquad \hat H =c\, {\mbox{\boldmath$\alpha$}}\cdot\textbf{p} + (\beta-1) c^2 +
V(\textbf{r},t) ,
\label{Time_Dirac1}
\end{equation}
where $\psi(\textbf{r},t)$ denotes the Dirac bispinor and $\vec{\alpha}$,
$\beta$ are the Dirac matrices. The two-center potential $V(\vec{r},t)$
consists of the nuclear Coulomb potentials of the target and
projectile
\begin{equation}
V(\vec{r},t)  = V^{A}_{\rm nucl}(r_A)+ V^{B}_{\rm nucl}(r_B) ,
\end{equation}
where indices $A$ and $B$ correspond to the target and projectile,
respectively.

Eq.~(\ref{Time_Dirac1}) is solved using the coupled-channel
approach with time-independent finite basis set $\{\varphi_k (\vec{r})\}$:
\begin{equation}
\psi(\vec{r},t) = \sum_k C_k(t) \, \varphi_k(\vec{r}) ,
\end{equation}
\begin{equation}
i \, S \, \frac{d \vec {C}(t)}{dt} = H(t) \,\vec {C}(t) .
\label{finite_bas1}
\end{equation}
Here $\vec{C}$ is the vector which incorporates the expansion coefficients
$C_k(t)$,
$H$ and $S$ are the Hamiltonian and overlapping matrices,
\begin{equation}
H_{k j} = \langle \varphi_k \mid \hat H \mid \varphi_j \rangle \,, \qquad
S_{k j} = \langle \varphi_k \mid \varphi_j \rangle .
\end{equation}
We note that in contrast to our previous work~\cite{Tup_10}, where the
time-dependent basis functions were employed, the differential matrix
equation~(\ref{finite_bas1}) has a simpler form. To solve 
equation~(\ref{finite_bas1}) we apply the
Crank-Nicolson (CN) method \cite{Crank_47, Varga_00}. 
In this method a short-time evolution operator, 
$\psi(t+\Delta t)=\hat U_{\rm CN}(t+\Delta t,t)\psi(t)$, is approximated by
\begin{equation}
\hat U_{\rm CN}(t+\Delta t,t) =
\left[ 1 + \frac{i \, \Delta t}{2} \,\, \hat H(t+\Delta t/2) \, \right ]^{-1} \, 
\left[ 1 - \frac{i \,\Delta t}{2} \, \hat H(t+\Delta t/2)  \right ].
\label{crank1}
\end{equation}
$\hat U_{\rm CN}$, being a unitary operator, conserves the norm of the wave
function. The CN method is known as stable and accurate  up to the $(\Delta t)^2$
terms included. 

With the CN method, the time-dependent equation~(\ref{finite_bas1}) can be
written as
\begin{equation}
\vec{C}(t+\Delta t) = U_{\rm CN}(t+\Delta t,t)~\vec{C}(t),
\label{crank2}
\end{equation}
where
\begin{equation}
U_{\rm CN}(t+\Delta t,t) =
\left[ 1 + \frac{i \, \Delta t}{2} \,S^{-1} \, H(t+\Delta t/2) \,
\right ]^{-1} \, 
\left[ 1 - \frac{i \,\Delta t}{2} \, S^{-1} \, H(t+\Delta t/2)  \right ].
\label{crank3}
\end{equation}
We emphasize that, in contrast to the operator $\hat U_{\rm CN}$, the matrix
$U_{\rm CN}$ is not unitary, since the matrices $S$
and $H$
do not
commute. However, the matrix $U_{\rm CN}$ also preserves the wave function norm
(see the Appendix)
\begin{equation}
\langle \psi(t+\Delta t) |  \psi(t+\Delta t) \rangle =
\vec{C}^{+}(t+\Delta t) \, S \, \vec{C}(t+\Delta t) =
\langle \psi(t) |  \psi(t) \rangle = 1 .
\end{equation}
To determine the coefficients $\vec{C}(t+\Delta t)$ at each time step 
we have to solve the following system of linear equations
\begin{equation}
\left [ S \, + \frac{i \Delta t}{2} \,\,  H \, \right ] \, 
\vec{C}(t+\Delta t) \,=\,
\left[ S - \frac{i \Delta t}{2}\, H \right ] \, \vec{C}(t).
\label{crank4}
\end{equation}
%
\subsection{Basis of cubic Hermite splines }
\label{sec:A2}
\begin{figure}
\centering 
\includegraphics[trim=0 0 0 0, clip, width = 0.5\textwidth]{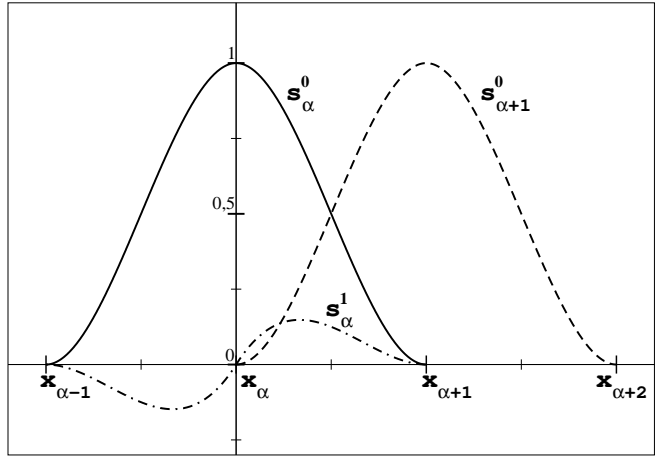}
\vspace{-2 mm}
\caption{The Hermite basis splines.}
\label{Fig_splines}
\end{figure}

%
In this paper we use a basis of piecewise Hermite cubic splines.
Let us  consider a partition of the interval $[a, b]$ into $N$ subintervals:  $a
= x_0 < x_1 < \ldots < x_N = b$, with the length
$h_\alpha = x_\alpha - x_{\alpha-1}$ of the $\alpha$-th interval.
We introduce two basis piecewise
functions $s^{0}_\alpha (x)$ and $s^{1}_\alpha (x)$ for each point $x_\alpha$
($\alpha=1,\ldots, N-1$)~\cite{De_Boor_01}:
\begin{equation}
s^{0}_\alpha(x)= \left \{
\begin{array}{lc}  \displaystyle
\frac{(x-x_{\alpha-1})^2}{h_\alpha^3} \, [2(x_\alpha-x)+h_\alpha] &
\displaystyle \qquad \quad x_{\alpha-1} \le x \le x_\alpha
\\[4mm] \displaystyle
\frac{(x_{\alpha+1}-x)^2}{h_{\alpha+1}^3} \, [2(x_{\alpha}-x)+h_{\alpha+1}] &
\displaystyle \qquad \quad x_\alpha \le x \le x_{\alpha+1}
\\[4mm] \displaystyle 0 & \qquad \quad \hbox{otherwise}
\end{array} \right .
\end{equation}
and 
\begin{equation}
s^{1}_\alpha(x)= \left \{
\begin{array}{lc}  \displaystyle
\frac{(x-x_{\alpha-1})^2}{h_{\alpha}^2} \, (x - x_\alpha) & \qquad \quad
\displaystyle x_{\alpha-1} \le x \le x_\alpha
\\[4mm] \displaystyle
\frac{(x_{\alpha+1}-x)^2}{h_{\alpha+1}^2} \, (x-x_\alpha) & \qquad \quad
\displaystyle x_\alpha \le x \le x_{\alpha+1}
\\[4mm] \displaystyle 0 & \qquad \quad \hbox{otherwise}
\end{array} \right ..
\end{equation}
The values of these functions and their first derivatives at the nodal points
are given by
\begin{equation}
s^{0}_\alpha(x_\beta) = \delta_{\alpha,\beta} \,, \qquad 
\frac{ds^{0}_\alpha}{dx} (x_\beta) = 0 \,, \qquad
s^{1}_\alpha(x_\beta) = 0 \,, \qquad \frac{ds^{1}_\alpha}{dx}(x_\beta) = 
\delta_{\alpha,\beta} \,.
\end{equation}
The functions $s^{0}_{\alpha}$ and  $s^{1}_{\alpha}$ are displayed in
Fig.~\ref{Fig_splines}. 
Both Hermite splines $s^{\mu}_{\alpha}$~($\mu=0,1$) are continuously
differentiable at all points, in contrast to the second derivatives, which are
discontinuous
at the points $x_{\alpha-1}$, $x_{\alpha}$, and $x_{\alpha+1}$. Since 
the $s^{\mu}_{\alpha}$ splines vanish outside the interval
($x_{\alpha-1},x_{\alpha+1}$), we can write
\begin{equation}
s^{\mu}_{\alpha}(x) \, s^{\nu}_{\beta}(x) =0 \qquad \hbox{if}
\qquad |\alpha-\beta| \ge 2 \,, \qquad (\mu, \nu =0,1) \,.
\end{equation}
Therefore the Hamiltonian and overlapping matrices in this basis are
sparse.

The Hermite cubic interpolation expansion for a function $f(x)$ 
with the boundary conditions $f(a)=f(b)=f^{\prime}(a)=f^{\prime}(b)=0$
is given by a rather simple equation:
\begin{equation}
f(x)= \sum_{\alpha=1}^{N-1} \left[ f(x_\alpha) \,s^{0}_\alpha(x) + 
f^{\prime}(x_\alpha) \,s^{1}_\alpha(x) \right] \,.
\label{splines_interpol}
\end{equation}
It should be noted that the piecewise Hermite cubic spline interpolation is
one of the best choice of the cubic spline interpolation
schemes~\cite{De_Boor_01}. 
\subsection{Monopole approximation (1D) }
\label{sec:A3}
In this subsection we consider the collision of
a bare nucleus (projectile) with
a H-like heavy ion (target) in the central-field
(monopole) approximation. 
Within this approximation only the monopole part of the reexpansion of the
projectile Coulomb potential at the target center is taken
into account:
\begin{equation}
V^B_{\rm mon}(r,t) =
\left  \{ \begin{array}{cc} \displaystyle
-\frac{Z}{r} & \qquad r \ge R(t)
\\[3mm] \displaystyle
-\frac{Z}{R(t)} & \qquad r < R(t)
\end{array} \right ..
\label{monopole1}
\end{equation}
Here $R(t)$ denotes the time-dependent distance between the target $A$
(at the rest) and the moving projectile $B$, and, for simplicity, the 
point-charge nuclear model is used.

We should stress that in our one-center monopole approximation
the electron-nucleus interaction potential is centered at the target $(A)$
position,
in contrast to the center of nuclear mass position, which was employed in
Refs.~\cite{Muller_72, Soff_78, Reus_84}.

In the central-field approximation the time-dependent wave function
$\psi(\vec{r},t)$  is the Dirac bispinor
\begin{equation}
\psi_{n\kappa m}(\vec{r},t) =
\left  ( \begin{array}{l} \displaystyle
\,\, \frac{~P_{n \kappa}(r,t)}{r} \,  \chi_{\kappa m}(\Omega)
\\[4mm] \displaystyle
i \, \frac{Q_{n \kappa}(r,t)}{r} \, \chi_{-\kappa m}(\Omega)
\end{array} \right ) \,,
\end{equation}
where $P_{n \kappa}(r,t)$ and $Q_{n \kappa}(r,t)$ are the large and small radial
components, respectively, 
$\chi_{\kappa m}(\Omega)$  is the spherical spinor,
and $\kappa=(-1)^{l+j+1/2}(j+1/2)$ is the
Dirac angular quantum number. 

The time-dependent radial Dirac equation can be
written in the form
\begin{equation}
\left  \{ \begin{array}{lll} \displaystyle
i \, \frac{\partial}{\partial t} \, P(r,t)  &=& \displaystyle
c \, \left[-\frac{d}{dr}+ \frac{\kappa}{r} \right] \, Q(r,t)+ 
\left [V^{A}(r)+V^{B}_{\rm mon}(r,t) \right] \, P(r,t)
\\[4mm] \displaystyle
i \, \frac{\partial}{\partial t} \, Q(r,t) &=&  \displaystyle
c \, \left[~~\frac{d}{dr}+ \frac{\kappa}{r} \right] P(r,t) +
\left[ V^{A}(r)+V^{B}_{\rm mon}(r,t) -2c^2 \right] \, Q(r,t) \,
\end{array} \right .,
\label{Dir1D}
\end{equation}
where $c$ is the speed of light. The functions $P(r,t)$ and $Q(r,t)$ are expanded in
the finite basis set
of cubic Hermite splines $s^{\mu}_\alpha(r)$ with the Hermite components
$(\mu=0,1)$
\begin{equation}
\left  \{ \begin{array}{lll} \displaystyle
P(r,t) &=&  \displaystyle  \sum_{\alpha, \mu} \, C^{P}_{\alpha \mu}(t)
\, s^{\mu}_{\alpha}(r)
\\[4mm]  \displaystyle
Q(r,t) &=&  \displaystyle  \sum_{\alpha, \mu} \, C^{Q}_{\alpha \mu}(t)
\, s^{\mu}_{\alpha}(r)
\end{array} \right ..
\label{expan1}
\end{equation}
Substituting this expansion into Eq.~(\ref{Dir1D}) leads to the
time-dependent matrix equation for the vector of coefficients $\vec{C}(t)$: 
\begin{equation}
i\frac{ \, \partial}{\partial t\,} \vec{C}(t) = S^{-1} \, H(t) \, \vec{C}(t)
\label{matC1d} \,, \qquad \vec{C}(t) = 
\left( \begin{array}{c}
\vec{C}^P(t)
\\[2mm]
\vec{C}^Q(t)
\end{array} \right) \,,
\end{equation}
where $S$ is the overlapping matrix and $H(t)$ is the matrix of the radial Dirac
operator $\hat H(t)$,
\begin{equation}
\hat H(t) = \left ( \begin{array}{cc} \displaystyle
V^{A}(r)+V^{B}_{\rm mon}(r,t) & \displaystyle
c \, \left[-\frac{d}{dr}+\frac{\kappa}{r} \right]
\\[4mm] \displaystyle
c \,\left[\frac{d}{dr}+\frac{\kappa}{r} \right] & \displaystyle
V^{A}(r)+V^{B}_{\rm mon}(r,t) -2c^2 \,
\end{array} \right ).
\label{oper1D}
\end{equation}
Eq.~(\ref{matC1d}) is solved by  the CN method
with $\vec{C}(t=-\infty)$
corresponding to the $1s$ state of the target H-like ion.

The monopole approximation is
rather crude at large internuclear distances and can not be applied to
investigation of the charge-transfer processes. 
However, the simple one-center calculations give some useful information about
electron-excitation and ionization processes, and can be used to estimate
the role of the negative-energy Dirac continuum.
%
\subsection{Axially-symmetric field (2D)} 
\label{sec:A4}
In this subsection we describe the 2D approach for heavy-ion
collisions. In this approach the process is approximated by the head-on
collision and the time-dependence of the internuclear distance is assumed to be
equal to $R(t)=\sqrt{Z^2(t)+b^2}$ (see Fig.~\ref{fig_geometry}). 

Since in the case under consideration the external field is axially
symmetric, the Dirac operator $\hat H$
commutes with the $z$-component $\hat J_z$ of the total angular
momentum of the electron. Then the wave function can be chosen as an
eigenfunction of the operator~$\hat J_z$,
\begin{equation}
\hat J_z \, \psi_m(\vec{r},t) = m \, \psi_m(\vec{r},t) 
\end{equation}
with a half-integer quantum number $m$.

In the cylindrical coordinates ($\rho,\phi,z$) the 4-component wave function
$\psi_m(\vec{r},t)$ can be written as
\begin{equation}
\psi_m(\rho, \phi, z, t)=
\frac{1}{\sqrt{2\pi}}\frac{1}{\sqrt{\rho}}
\begin{pmatrix} 
U^1_m(\rho,z,t)\exp[i(m-\frac{1}{2})\phi] \\[0.5mm]
U^2_m(\rho,z,t)\exp [i(m+\frac{1}{2})\phi] \\[0.5mm]
iU^3_m(\rho,z,t)\exp[i(m-\frac{1}{2})\phi]\\[0.5mm]
iU^4_m(\rho,z,t)\exp[i(m+\frac{1}{2})\phi]
\end{pmatrix}.
\label{subst}
\end{equation}
The factor $\sqrt{\rho}$ is introduced in the definition of the
wave function $U_m(\rho,z,t)$ \cite{Schluter_83} in order to simplify
the integration over the variable $\rho$ in the matrix elements of the Dirac
and unit operators. The normalization condition of the wave function
is given by
\begin{equation}
\langle \psi_m | \psi_m \rangle = \int \limits_{-\infty}^{\infty} dz \,
\int \limits_{0}^{\infty} d\rho \, U^{+}_m(\rho,z,t) \, U_m(\rho,z,t) = 1 \,.
\end{equation}
Substituting Eq.~(\ref{subst}) into the Dirac
equation~(\ref{Time_Dirac1}),
we obtain 
\begin{equation}
i \frac{\partial}{\partial t} \, U_m(\rho,z,t) = \hat H_{\rm C} \,
U_m(\rho,z,t) ,
\label{Time_Dirac2}
\end{equation}
where the Hermitian operator $\hat H_{\rm C}$ is the cylindrical part of the
Dirac operator
\begin{equation}
\hat H_{\rm C} =
\begin{pmatrix} \displaystyle
\quad V \quad & \displaystyle 0 & \displaystyle c\frac{\partial}{\partial z}
& \displaystyle \hat{K}\\[0.8mm]
0 &  \displaystyle \quad V \quad  & \displaystyle -\hat{K}^{+} &
\displaystyle -c\frac{\partial}{\partial z}\\[0.8mm]
 \displaystyle -c\frac{\partial}{\partial z} & \displaystyle -\hat{K}
& \displaystyle \quad V-2c^2 \quad & \displaystyle 0\\[0.8mm]
-\hat{K}^{+} &  \displaystyle c\frac{\partial}{\partial z} & \displaystyle 0
& \displaystyle \quad V-2c^2 \quad 
\end{pmatrix} ,
\label{bigmatrixfinal}
\end{equation}
\begin{equation}
\hat{K}=c\left(\frac{\partial}{\partial \rho}+\frac{m}{\rho} \right)\,.
\end{equation}
We solve the time-dependent Dirac equation~(\ref{Time_Dirac2}),
using the finite basis expansion in both $\rho$ and $z$ variables
\begin{equation}
U^k(\rho,z,t) = \sum_{\alpha, \mu} \sum_{\beta, \nu} \,
C^{k}_{\alpha \mu, \beta \nu}(t) \, s^{\mu}_\alpha(\rho) \,
s^{\nu}_\beta(z) \,.
\end{equation}
Here index $k=1,2,3,4$ enumerates the $U(\rho,z,t)$ components.
The coefficients $\vec{C}(t)$ can be found solving the
time-dependent matrix equation~(\ref{finite_bas1}).
%
\subsection{ Full 3D approach }
\label{sec:A5}
%
In the 3D case we use the Cartesian coordinates ($x,y,z$) and the finite basis
spline expansion of the wave function $\psi(\vec{r},t)$ in the form
\begin{equation}
\psi^k(x,y,z,t) = \sum_{\alpha, \mu}  \sum_{\beta, \nu} \, 
\sum_{\gamma, \lambda} \, C^{k}_{\alpha \mu, \beta \nu, \gamma \lambda }(t) \,
s^{\mu}_\alpha(x) \, s^{\nu}_\beta(y) \, s^{\lambda}_\gamma(z) \,.
\end{equation}
The index $k=1,2,3,4$ enumerates the components of the Dirac
bispinor, indices $\mu,\nu, \lambda=0,1$ denote the type of the cubic Hermite
splines,
and indices $\alpha, \beta, \gamma$ label the splines, centered at the
different
points. The total number of the basis functions is equal to 
$N=4 \cdot 2N_x \cdot 2N_y \cdot 2N_z$, where $N_x$, $N_y$ and $N_z$
are the numbers of grid points in the $x$, $y$, and $z$ directions,
respectively.

In the 3D case we are faced with the huge sparse overlapping $S$ and Hamiltonian
$H(t)$
matrices. 
It is possible to store only nonzero elements of these matrices in computer memory.
Solving the time-dependent equation by the CN method~(\ref{crank4})
we have to calculate the inverse $S^{-1}$ matrix. A fast algorithm based on
factorization of the overlapping matrix into Kronecker's (direct) matrix
production 
$S = S_x \otimes S_y \otimes S_z$ and the inverse matrix calculation 
\cite{Lancaster_85}
$S^{-1} = S^{-1}_x \otimes S^{-1}_y \otimes S^{-1}_z$ are used.

Eq.~(\ref{crank4}) is solved by the stabilized biconjugate
gradient method~\cite{Saad_03}. The generalization
of this method for the case of
complex matrices is given in Ref.~\cite{Joly_93}.

The coefficients $\vec{C}$ at the initial time point can be determined
using the interpolation properties of the cubic Hermite splines
(\ref{splines_interpol})
\begin{equation}
C^{k}_{\alpha \mu, \beta \nu, \gamma \lambda }= \left .
\left( \frac{\partial}{\partial x} \right)^{\mu} \, 
\left( \frac{\partial}{\partial y} \right)^{\nu} \, 
\left( \frac{\partial}{\partial z} \right)^{\lambda} \, 
\psi_0^k(x,y,z) \right |_{x_\alpha, y_\beta, z_\gamma} \,, \qquad 
\mu, \nu, \lambda=0,1 \,,
\end{equation}
where $\psi_0(x,y,z)$ is the ground state wave function of the H-like
ion in the central-field approximation.
%
\subsection{Transition amplitudes. Contribution of the negative-energy
Dirac continuum}
\label{sec:A6}
%
We consider here, for simplicity, one-center transitions
between the target states, assuming that the target is at rest.
For the one-electron system the transition amplitude is defined by
\begin{align}
&T_{ji} = 
\langle \psi_j^{(0)}(t) | \psi_{i}(t) \rangle, \quad t\rightarrow\infty,
\label{A6:1}
\end{align}
where $\psi_{j}^{(0)} (\vec r,t)=e^{-i\varepsilon_j t}\phi_j(\vec r)$
is a one-electron stationary wave function of the unperturbed target
Hamiltonian ($\hat{H}^{(0)} \phi_j = \varepsilon_j \phi_j$) and 
$\psi_i (\vec r,t)$ is the wave function of the colliding system with the
initial condition:
\begin{align}
\psi_i(\vec{r},t) \rightarrow
e^{-i\varepsilon_i t} \phi_{i}(\vec{r}),
\quad t \rightarrow -\infty.
\label{A6:2}
\end{align}
The corresponding probability is equal to
$P_{ji} = |T_{ji}|^2$.
In particular, the probability to find the
electron after the collision in the ground $1s$ state of the target is
\begin{equation}
P_{1s} =  |\langle \psi^{(0)}_{1s}(t) |
\psi_{1s}(t) \rangle|^2, \quad
t\rightarrow\infty.
\label{A6:3}
\end{equation}
Formally, we can also define the total transition probability $P^{(-)}$ to the
negative-energy states
\begin{equation}
P^{(-)} = \sum_{j}^{\varepsilon_j \leq -2c^2}  |\langle \psi_{j}^{(0)}(t)
|
\psi_{1s}(t) \rangle|^2, \quad t\rightarrow\infty.
\end{equation}
To be closer to the real situation, we should consider the many-electron
picture, where all the negative-energy continuum states are occupied by electrons
according to the Pauli principle.
Then, neglecting the electron-electron interaction,
the one-electron wave functions $\psi^{(0)}_{1s}(\vec r,t)$ and
$\psi_{1s}(\vec r,t)$ in Eq.~(\ref{A6:3})
have to be replaced by the Slater determinants
\begin{equation}
\Psi^{(0)}_{1s}(\vec r_1,\ldots,\vec r_{N_e},t) =
\frac{1}{\sqrt{N_e!}} \, \det
\{\psi_k^{(0)}(\vec{r}_l,t)\} \,,
\quad
\Psi_{1s}(\vec r_1,\ldots,\vec r_{N_e},t)= 
\frac{1}{\sqrt{N_e!}} \, \det
\{\psi_k(\vec{r}_l,t)\}
\,,
\end{equation}
where $N_e$ is the number of electrons, that
includes the one $1s$ and all negative-energy continuum electrons. Then, the
corrected probability $\overline P_{1s}$ to find the system after the collision
in the ground $1s$ state is given by
\begin{equation}
\overline P_{1s}=
|\langle \Psi_{1s}^{(0)}(t)|\Psi_{1s}(t)\rangle|^2  
=
|\det\{ \langle \psi_k^{(0)}(t) |
\psi_l(t) \rangle \} |^2 \,,
\label{correct_prob}
\end{equation}
where $k$ and $l$ run over the one $1s$ and all negative-energy continuum
states. In deriving Eq.~(\ref{correct_prob}) we have used the fact that the scalar
product of two Slater determinants is equal to the determinant of the scalar
product of the one-electron wave functions~\cite{McWeeny_01}.

The calculation of the corrected probability $\overline P_{1s}$ is
much more time consuming than the calculation of $P_{1s}$. In the
present paper we performed this calculation
in the 1D case only.

We note that the same result, which is given by
Eq.~(\ref{correct_prob}), can be obtained using the second quantization
formalism~\cite{Fradkin_91}.
\section{RESULTS OF THE CALCULATIONS AND DISCUSSION}
\label{sec:results}
%
In this section we present the results of the calculations within the monopole
(1D) and axially-symmetric (2D) approximations, and the full 3D approach.
We consider the straight-line collision of the 
H-like uranium (target, $A$) being initially in the ground state with the
bare uranium nucleus (projectile, $B$) at the 6 MeV/u energy and the impact
parameter~$b$ (see Fig.~\ref{fig_geometry}).
We choose the coordinate system with the origin at the center of the fixed
target
and the $z$-axis parallel to the straight-line trajectory of the projectile.
Unless stated otherwise, the model of the nuclear charge distribution employed
is a uniformly
charged sphere of radius $R_{\rm n}=\sqrt{5/3}R_{\rm RMS}$, where $R_{\rm
RMS}$ is the root-mean-square nuclear radius. 
Then the Coulomb potential of the nucleus is given by
\begin{equation}
V_{\rm nucl}(r) =
\left  \{ \begin{array}{cc} \displaystyle
- \, \frac{Z}{r} & \qquad r \ge R_n
\\[3mm] \displaystyle
- \frac{Z}{2R_n} \, \left(3-\frac{r^2}{R_n^2} \right) & \qquad r < R_n \,
\end{array} \right ..
\end{equation}
According to Ref.~{\cite{Kozhedub_08}},
we use $R_{\rm RMS}=5.8569(33)$~fm
for the uranium nuclear radius.
%
\subsection{Monopole approximation (1D)} 
In the monopole approximation we use the basis set of $384$ splines ($96$
grid points) that is sufficient to obtain the results with a high accuracy.
This can be seen from Table~\ref{1D_compar}, where we compare our
data for the energy of the $1s$ state of H-like ions,
calculated for the point-charge nucleus in the finite basis approximation,
with the exact analytical values.

The semi-logarithmic grid $\zeta_\alpha =\eta \,  r_\alpha +\xi \, {\rm
ln}(r_\alpha) $, 
proposed by Brattsev~\cite{Brattsev_66}
and widely used in nonrelativistic~\cite{Chernysheva_66} and
relativistic~\cite{Brattsev_77}
atomic calculations, is employed to generate the set of points
$r_{\alpha}$
(points $\zeta_\alpha$ are taken with a constant step).
\begin{table}[ht]
\caption{The $1s$ state energy of H-like ions (in a.u.) for the point-charge
nucleus.}
\label{1D_compar}
\begin{center}
\begin{tabular}{|c|c|c|}
\hline
Z   & \hspace{2mm} Finite basis \hspace{2mm} & \hspace{6mm} Exact values
\hspace{6mm} \\
\hline
92  & -4861.1979 & -4861.1979 \\[0mm]
100 & -5939.1952 & -5939.1952 \\[0mm]
130 & -12838.926 & -12838.920 \\[0mm]
\hline
\end{tabular}
\end{center}
\end{table}

In the monopole approximation we can not calculate the charge transfer
probabilities. However, we can evaluate the $1s$ state target population
probability
$P_{1s}(b)$ (probability to stay in the $1s$ target state after the collision)
and
the transition probability to the negative-energy continuum states $P^{(-)}(b)$
as functions of the impact parameter $b$. 
The collision was considered in a spherical box, with the target placed
at the box
center. The radius of the box was taken to be $19/Z {\rm ~a.u.} \simeq
10924$~fm. This value should be compared with the mean
radius of the $1s$ orbital of H-like uranium equal to $\langle r
\rangle \simeq 0.0135$~a.u.$\simeq713$~fm.

\begin{table}[htb]
\caption{The average energy $E_{\rm min}(b)$ at the minimal internuclear
distance ($R=b$), as a function of the impact parameter $b$ (in fm).
The calculations are performed for the point-like and uniformly charged
sphere nuclear models ($R_{\rm RMS}=5.8569$~fm).
The monopole (1D) approximation is used.
}
\label{1D_collision1}
\begin{center}
\begin{tabular}{|c|c|c|}
\hline
\multicolumn{1}{|c }{} & \multicolumn{2}{ c|}{ $E_{\rm min}/mc^2+1$} \\[0mm]
\hline
~~$b$~~ & ~~Point-charge nucleus~~ &  ~~Sphere nuclear model~~ \\[1mm]
\hline
15  &  ~~-1.353~~     &  ~~-1.224~~  \\[0mm]
20  &    -1.124       &    -1.046    \\[0mm]
25  &    -0.967       &    -0.913    \\[0mm]
30  &    -0.849       &    -0.810    \\[0mm]
40  &    -0.680       &    -0.656    \\[0mm]
50  &    -0.561       &    -0.545    \\[0mm]
\hline
\end{tabular}
\end{center}
\end{table}

In Table~\ref{1D_collision1} we present the results of our calculations
for the minimal energy $E_{\rm min}(b)$,
as a function of the impact parameter $b$.
The minimal energy $E_{\rm min}(b)$ was calculated as the expectation value of
the time-dependent Hamiltonian $\hat H(t)$ (see~Eq.~(\ref{oper1D})) at
the shortest internuclear distance $R(0)=b$, which corresponds to $t=0$.
As one can see from Table~\ref{1D_collision1}, the average energy $E_{\rm min}(t,b)$
dives into the negative-energy continuum at the impact parameter slightly
bigger than $b=20$~fm. It should be noticed that the stationary $1s$ state of the
quasi-molecule calculated within the 1D model for the point-charge nucleus
dives into the negative-energy continuum at the critical internuclear distance 
$R_{\rm cr}=25.5$~fm~\cite{Tup_10}.
%

\begin{table}[htb]
\caption{The population probability of the $1s$ target state $P_{1s}(b)$
and the negative-energy continuum population probability $P^{(-)}(b)$
at the infinite time limit ($t \to \infty$) as functions of the impact
parameter $b$ (in fm). The monopole (1D) approximation is used.
}
\label{1D_collision2}
\begin{center}
\begin{tabular}{|c|c|c|c|c|}
\hline
\multicolumn{1}{|c|}{} & \multicolumn{2}{|c|}{Point-charge nucleus} &
\multicolumn{2}{|c|}{Sphere nuclear model}\\[0mm]
\hline
~~~~$b$~~~~ & \hspace{7mm}  $P_{1s}$  \hspace{7mm} &
\hspace{7mm} $P^{(-)}$ \hspace{7mm} & \hspace{7mm}  $P_{1s}$  \hspace{7mm} &
\hspace{7mm} $P^{(-)}$  \hspace{7mm} \\[1mm]
\hline
15  &   0.549435  &  ~~4.70 $\times$ $10^{-3}$~~ & 0.610272 &  3.43 $\times$ $10^{-3}$~~ \\[0mm]
20  &   0.669281  &    2.59 $\times$ $10^{-3}$   & 0.706189 &  2.00 $\times$ $10^{-3}$~~ \\[0mm]
30  &   0.811566  &    0.87 $\times$ $10^{-3}$   & 0.826959 &  0.72 $\times$ $10^{-3}$~~ \\[0mm]
40  &   0.886131  &    0.32 $\times$ $10^{-3}$   & 0.893379 &  0.27 $\times$ $10^{-3}$~~ \\[0mm]
50  &   0.928079  &    0.12 $\times$ $10^{-3}$   & 0.931794 &  0.11 $\times$ $10^{-3}$~~ \\[0mm]
\hline
\end{tabular}
\end{center}
\end{table}
%

In Table~\ref{1D_collision2} we present the results of our calculations
for  the $1s$ population probability $P_{1s}(b)$ and the
negative-energy continuum population probability $P^{(-)}(b)$
as functions of the impact parameter $b$.
The $1s$ population probability $P_{1s}(b)$ was
calculated in the one-electron picture. 
%
\begin{table}[htb]
\caption{The population probability of the
$1s$ target state calculated within the one-electron $\left(P_{1s}(b)\right)$  and 
many-electron $\left(\overline P_{1s}(b)\right)$ pictures as a function of the
impact parameter $b$ (in fm).
 The calculations are performed for the point-like 
and uniformly charged sphere nuclear models ( $R_{\rm RMS}=5.8569$~fm).
The monopole (1D) approximation is used.
}
\label{1D_collision3}
\begin{center}
\begin{tabular}{|c|c|c|c|c|}
\hline
\multicolumn{1}{|c|}{} & \multicolumn{2}{|c|}{Point-charge nucleus} &
\multicolumn{2}{|c|}{Sphere nuclear model}\\[0mm]
\hline
~~$b$~~ &
\hspace{4mm}  $\overline P_{1s}$         \hspace{4mm} &
\hspace{3mm}  $\overline P_{1s}-P_{1s}$  \hspace{3mm} &
\hspace{4mm}  $\overline P_{1s}$         \hspace{4mm} & 
\hspace{2mm}  $\overline P_{1s}-P_{1s}$  \hspace{2mm} \\
\hline
%
 15 & ~~0.550244~~ & ~~8.09 $\times$ $10^{-4}$~~  &~~~ 0.610755 ~~~ &  4.84 $\times$ $10^{-4} $
\\[0mm]
 20 & 0.669606     &    3.25 $\times$ $10^{-4}$  &    0.706402     &  2.14 $\times$ $10^{-4} $
\\[0mm]
 30 & 0.811627     &    0.61 $\times$ $10^{-4}$  &    0.827004     &  0.45 $\times$ $10^{-4} $
\\[0mm]
 40 & 0.886144     &    0.13 $\times$ $10^{-4}$  &    0.893389     &  0.11 $\times$ $10^{-4} $
\\[0mm]
 50 & 0.909947     &    0.03 $\times$ $10^{-4}$  &    0.931796    &  0.03 $\times$ $10^{-4} $
\\[0mm]
\hline
\end{tabular}
\end{center}
\end{table}

We also calculated the corrected $1s$ population probability
$\overline P_{1s}(b)$ within the many-electron picture, described in
Sec~\ref{sec:A6}. The calculations have been performed for the
point-like and uniformly charged sphere nuclear models.
The obtained values are presented in Table~\ref{1D_collision3}.
The comparison of the $\overline P_{1s}(b)$ data with the values
obtained in the one-electron picture shows
that the role of the negative-energy continuum is rather
small, and a larger effect comes from the nuclear charge distribution.
%
%
%

The high probability of staying in the initial state
of the target can be explained by the fact that the velocity of the
incident particle
$v_{\rm p}$ is much smaller than the velocity of electron motion $v_{\rm e}$
in the nuclear field, $v_{\rm p}/v_{\rm e} \sim 0.16$.
%
\subsection{Axial-symmetry approximation (2D)} 
In the axial-symmetry approximation the calculations were performed
using 200 splines (100 grid points) along the $z$ axis and
52 splines (26 grid points) for the variable $\rho$ on the uniform grid. Thus,
the total
number of the basis functions was equal to $4 \times 200 \times 52 =  41600$.
The cylindrical box $20000 \times 5000$~fm$^2$ was used, and the target position
was shifted by $5000$~fm from the box center along the $z$
axis in the direction of the initial projectile position.
This was done to minimize the influence of the box borders on the
time-dependent wave function after the collision.
The number of time steps, which was used to solve the time-dependent equation,
was equal to $15000$. The point-charge nuclear model was used for both colliding nuclei.
The initial $1s$ wave function, localized at the target
ion,
was calculated as an eigenfunction of the Hamiltonian matrix in the same
basis set. The obtained energy value, $-4849$~a.u., is close to
the exact one, which is equal to $-4861$~a.u. 

\begin{figure}[ht]
\centering
\includegraphics[trim=0 0 0 0, clip, width = 0.7\textwidth]
{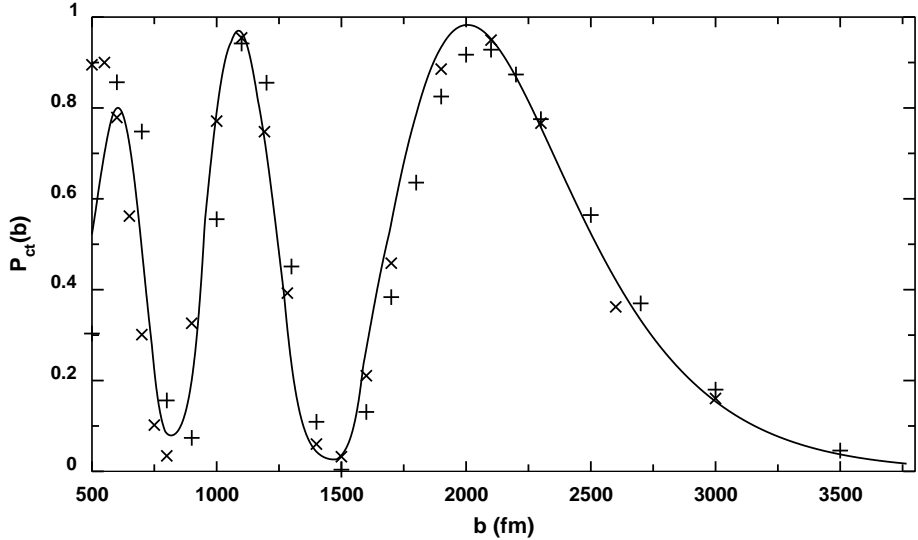} 
\caption{The charge-transfer probability $P_{\rm ct}(b)$ as a function
of the impact parameter $b$. The signs "+" and "$\times$"
 indicate the 2D and 3D results,  respectively. The solid line shows the results
from Ref.~\cite{Tup_10}.}
\label{capture}          
\end{figure}

The charge-transfer probability $P_{\rm ct}(b)$ was
calculated by
dividing the entire space into two equal parts and integrating the final
electron density over the part with the projectile.
The values of $P_{\rm ct}(b)$ obtained in the 2D approximation and in the
full 3D approach (the corresponding details are given in the next subsection),
and also the related data from Ref.~\cite{Tup_10} are presented in
Fig.~\ref{capture}. 
The evaluation was done within the one-electron picture.
One can observe a rather good agreement between the 2D and full 3D results
for the charge-transfer probability.
\subsection{Full 3D approach}
In the 3D case the charge-transfer probability was calculated using 
$40 \times 40 \times 80$ splines for each component of the four component
relativistic wave function on the 3D $(x,y,z)$ uniform space grid. The
total size of the finite basis set was equal to $512000$. The
point-charge nuclear model was used for both colliding nuclei. The
U$^{91+}$($1s$) initial state energy calculated with this basis is equal to
$-4711$~a.u., that fairly agrees with the exact energy value.

The rectangular box $6900 \times 6900 \times 13800$ fm$^3$ was used in
the 3D calculations and, as in the 2D case, the target position was shifted
from the box center along the $z$ axis in the direction of the initial
projectile position by the value equal to a quarter of the box length.
The time-dependent equation was solved using the CN method with 1024  time
steps. Again, the one-electron picture was
used.

The values of the charge-transfer probability $P_{\rm ct}(b)$
obtained with the 3D approach are shown in Fig.~\ref{capture}.
The results are in good agreement with the data obtained in
Ref.~\cite{Tup_10}.

%
%
%
\section{CONCLUSION}
%
In this paper we presented a new method for the relativistic calculations of
one-electron two-center quasi-molecular systems in both stationary and
time-dependent regimes using the finite basis set of cubic Hermite
splines. The calculations were performed for the low-energy
U$^{91+}(1s)$--U$^{92+}$
collision at the projectile energy $6$~MeV/u within the 1D, 2D and 3D
approaches.

In the 1D approximation we examined the influence of the
negative-energy Dirac continuum on the $1s$ population
probability. It was found that this
influence is rather small. It
should also be noted that the probability to find the electron in the $1s$
state of the target after the collision, calculated in the monopole
approximation, is quite large. This shows the adiabatic nature of the collision
process.

The charge-transfer probabilities were evaluated in the 2D
approximation and in the full 3D approach using
the one-electron picture. The results of the calculations are in
a good
agreement with each other, that also indicates the adiabatic nature of
the collision process. The obtained results agree also with 
our previous calculations performed by the Dirac-Sturm
method~\cite{Tup_10}.
\section*{\large Acknowledgments}
We thank S.~Hagmann and C.~Kozhuharov
for many helpful discussions.
This work was supported by RFBR
(Grants No. 13-02-00630 and  No. 11-02-00943-a), 
by the Ministry of Education and Science of the Russian Federation (Grant No. 8420),
by GSI, by DAAD, and by
the grant of the President of the Russian Federation (Grant No.
MK-2106.2012.2). 
The work of I.A.M. was also supported by the Dynasty foundation.
I.A.M., A.I.B., and Y.S.K. acknowledge financial support by the FAIR--Russia Research 
Center.
%
%
\clearpage
\section* {\large Appendix: Crank-Nicolson method for the
finite basis time-dependent equation in the non-orthogonal basis}
\label{sec:CN}
\setcounter{equation}{0}
\renewcommand{\theequation}{A\arabic{equation}}
%
Consider the time-dependent Dirac equation in the finite basis set
\begin{equation}
i \, S \, \frac{d \vec {C}^{i}(t)}{dt} = H(t) \,\vec {C}^{i}(t) \,,
\quad
\psi_{i}(\vec {r},t)=\sum_k C^{i}_{k}(t) \varphi_k (\vec {r}).
\label{finite_bas2}
\end{equation}
Here the index $i$ enumerates different solutions of the
time-dependent equation. In the Crank-Nicolson approximation
the coefficients $\vec{C}^i(t+\Delta t)$ can be determined from the
coefficients $\vec{C}^i(t)$
by solving the system of linear equations
\begin{equation}
\left [ S \, + \frac{i \Delta t}{2} \,\,  H \, \right ] \, 
\vec{C}^i(t+\Delta t) \,=\,
\left[ S - \frac{i \Delta t}{2}\, H \right ] \, \vec{C}^i(t) \,.
\label{crank5}
\end{equation}
Let us rewrite this equation in the following way
\begin{equation}
S^{1/2} \, \left [ 1 \, + \frac{i \Delta t}{2} \,\,  H^{(L)} \, \right ] \,
S^{1/2}
\vec{C}^i(t+\Delta t) =
S^{1/2} \, \left[ 1 - \frac{i \Delta t}{2}\, H^{(L)} \right ] \,
S^{1/2} \vec{C}^i(t),
\label{crank6}
\end{equation}
where $H^{(L)}$ is the Hamiltonian matrix in the L\"owdin representation
\cite{Lowdin_50}
$$
H^{(L)} \equiv S^{-1/2} \, H \,  S^{-1/2} \,.
$$
Eq.~(\ref{crank6}) is conveniently written in the form
\begin{equation}
\vec{C}^i(t+\Delta t) = U_{\rm CN} \, \vec{C}^i(t) \,,
\end{equation}
where the matrix $U_{\rm CN}$ is given by
\begin{equation}
U_{\rm CN} \equiv  S^{-1/2} \, (V^{-1})^+ \, V \, S^{1/2} \,,
\label{dirac_time11}
\end{equation}
with
\begin{equation}
V \equiv \left[ 1 - \frac{i \Delta t}{2}\, H^{(L)} \right ] \,.
\end{equation}
The matrices $V$ and $V^{+}$ commute, since $H^{(L)}$ is an Hermitian matrix.
We obtain
$$
U_{\rm CN} \,S \, U_{\rm CN}^{+} = 
\left (S^{1/2} \, V^{-1^+} \, V \, S^{-1/2} \right) \, S \,
\left (S^{-1/2} \, V^{+} \,V^{-1} \, S^{1/2} \right) = S \,
$$
and
\begin{equation}
\vec{C}^{i^{+}}(t+\Delta t) \, S \, \vec{C}^{j}(t+\Delta t) =
\vec{C}^{i^{+}}(t) \, U_{\rm CN}^{+} \, S \,  U_{\rm CN} \,
\vec{C}^{j}(t) =
\vec{C}^{i^{+}}(t) \, S \,  \vec{C}^{j}(t) \,.
\end{equation}
Thus, the Crank-Nicolson approximation preserves the norm of the wave function 
\begin{equation}
\langle \psi_{i}(t+\Delta t) |  \psi_{j}(t+\Delta t) \rangle =
\vec{C}^{i^{+}}(t+\Delta t) \, S \, \vec{C}^{j}(t+\Delta t) =
\langle \psi_{i}(t) |  \psi_{j}(t) \rangle \,.
\end{equation}
\pagebreak

\pagebreak
%
\end {document}